\documentclass[aps,prb,twocolumn]{revtex4}
\usepackage{tabularx,graphicx}
\usepackage{amsmath, amsthm, amssymb} 
\usepackage{bbm}

\def\be{\begin{equation}}
\def\ee{\end{equation}}
\def\bea{\begin{eqnarray}}
\def\eea{\end{eqnarray}}
\def\ba{\begin{array}}
\def\ea{\end{array}}

\begin{document}

\title{de Haas-van Alphen oscillations for non-relativistic fermions coupled to an emergent $U(1)$ gauge field}

\author{Lars Fritz}
\affiliation{Department of Physics, Harvard University, Cambridge MA 02138, USA}
\affiliation{Institut f\"ur Theoretische Physik,
Universit\"at K\"oln, Z\"ulpicher Stra\ss e 77, 50937 K\"oln, Germany}
\author{Subir Sachdev}
\affiliation{Department of Physics, Harvard University, Cambridge MA 02138, USA}

\date{\today}

\begin{abstract}
We investigate magento-oscillations in the specific heat of 
non-relativistic fermions with a Fermi surface minimally coupled to a fluctuating U(1) gauge field. 
Our study is motivated by the recent observation of quantum oscillations in the underdoped cuprates,
and by theoretical models of pocket Fermi surfaces realizing a non-Fermi liquid `algebraic charge liquid'.
Our main result is the computation of the order $1/N$ correction to the Lifshitz-Kosevich expression for
the oscillation amplitude in the dirty limit in a model with $N$ species of fermions.
\end{abstract}

\maketitle


\maketitle

\section{Introduction}\label{intro}
The problem of two dimensional non-relativistic fermions coupled to a gauge field has been intensively studied over the last couple of years in a variety of different contexts. It appears as a low energy description of different models of strongly correlated electronic systems, such as electrons in the fractional quantum Hall regime and in theories of non-Fermi liquid phases for the
underdoped cuprates.

Our study is motivated by the remarkable recent observations \cite{Doiron,Yelland,Bangura,Jaudet,Sebastian,LeBoeuf} 
of quantum oscillations in the underdoped cuprate superconductors at high magnetic fields. So far, these observations have been consistently interpreted using 
Fermi liquid models of the formation of Fermi pockets. The Fermi liquid theory of quantum oscillations \cite{Shoenberg} yields periodic behavior as a function of $1/B$, where $B$ is the applied magnetic field, with an amplitude given by the Lifshitz-Kosevich (LK) prefactor \cite{Lifshitz}.

However, as the precision and range of the observations increase, it would be useful to have theoretical
predictions for other candidate metallic ground states of the underdoped cuprates. To this end, we will examine
the amplitude of the quantum oscillations in `algebraic charge liquids' (ACL)\cite{SachdevDresden,Galitski,Moon}. The charged excitations in these
states are described by Fermi surfaces of spinless electrons coupled to an emergent U(1) gauge field, as we will
review in more detail in Section~\ref{sec:cuprates}. We will find that these systems also exhibit oscillations
which are periodic in $1/B$, with a prefactor with small but detectable deviations from the LK theory.

Ref.~\cite{Stamp} has also addressed this problem in the clean limit. However, they do not include the oscillatory terms in the gauge field propagator, which are responsible for the main effects we describe below.

It is also possible that our results have implications for quantum Hall systems, where similar theories apply
to compressible states at even denominator fillings. However, we will not explore this connection further here.

Our main analysis is a computation of the free energy of a system of $N$ species of fermions coupled to a fluctuating,
emergent U(1) gauge field. We compute the de Haas-van Alphen oscillations in the presence of an applied magnetic
field in the dirty limit to order $1/N$. The $N=\infty$ theory co-incides with the LK result. Our main results for the 
order $1/N$ corrections are shown in Eq.~\eqref{cvelectrons}, Eq.~\eqref{gaugenonosc}, and Eq.~\eqref{gaugeosc}. Most interestingly, we find a qualitative difference in the behavior  of the oscillations as a function of $T/\omega_c$, {\it i.e.} temperature over the cyclotron frequency, compared to Fermi liquid theory~\cite{Bondarenko, McCombe,Sullivan}. This comparison is plotted in Fig.~\ref{fig1}.
Additionally, for the specific heat, $c_V$, we find that the gauge field correction to the oscillatory term
has a temperature ($T$) dependence $\sim T \ln (1/T)$ which differs from the $\sim T$ dependence in the LK term,
and so may be experimentally detectable in recent and future experiments~\cite{Riggs}. 

The organization of the paper is as follows. In Section~\ref{sec:model} we introduce the generic model of fermionic degrees of freedom in the presence of weak scalar disorder coupled to an internal gauge degree of freedom. In Section~\ref{LL} we discuss the Landau level structure, and the role of disorder together with its treatment on a phenomenological level. We conclude this section by deriving the effective theory for the gauge field in the presence of disorder and Landau levels in Section~\ref{polarization}.

In Section~\ref{oscillatorygauge} we present a formalism to extract the oscillatory part of the gauge field contribution to order $1/N$. To this end we have to calculate oscillatory thermodynamic and transport quantities, part of which is outlined in Appendix~\ref{App:grandpotential}. In Section~\ref{freespec} we show how to calculate the specific heat of the entire coupled system to order $1/N$. In Section~\ref{specelec} we review the derivation of the specific heat and the associated de Haas-van Alphen oscillations of a disordered gas of electrons subject to a magnetic field. In Section~\ref{specgauge} we calculate the specific heat of the gauge field. In a first step (Section~\ref{nonosc}) we derive the specific heat of the non-oscillatory part, whereas in Section~\ref{oneelec} we calculate the oscillatory correction. In Section~\ref{sec:cuprates} we explain the meaning of our results in the context of a very recent spin density wave (SDW)+ gauge field description of the cuprates. Finally, we summarize in Section~\ref{summary}.

\section{Model}\label{sec:model}
The effective model which forms the basis for this paper was introduced in different contexts and its main characteristics are reviewed here. We consider a conventional Fermi gas with quadratic dispersion, which is minimally coupled to an internal $U(1)$ gauge field. The Lagrangian for this system generically reads
\begin{eqnarray}\label{eq:ferm1}
\mathcal{L}_f = \overline{f} \left( \partial_\tau-iA_\tau-\frac{(\nabla-i{\bf{A}})^2}{2m}-\mu\right)f \;,
\end{eqnarray}
where ${\bf A}$ denotes the internal $U(1)$ gauge field.
\subsection{Disorder broadened Landau levels in the dirty limit}\label{LL}
In our problem we consider the above system subject to an external perpendicular magnetic field ${\bf{B}}={\bf{\nabla}}\times {\bf{a}}$, which implies the fermions also couple minimally to the external gauge field ${\bf{a}}$ and causes the electronic system to organize its spectrum into Landau levels. In the absence of the internal $U(1)$ gauge field ${\bf {A}}$ we can diagonalize the electronic part and cast it as
\begin{eqnarray}\label{eq:ferm1mag}
\mathcal{L}_f = \overline{f} \left( \partial_\tau-\omega_{c}\left(n+\frac{1}{2}\right)\right)f \;, \nonumber \\ 
\end{eqnarray}
where $\omega_{c}={eB}/{m}$ is the cyclotron frequency, $n$ the Landau level, and the degeneracy of the Landau levels is given by
\begin{eqnarray}
\frac{1}{2\pi l_{B}^2}=\frac{m}{2\pi}\omega_{c}=\nu_0 \omega_{c}\;,
\end{eqnarray} 
where $\nu_0=\frac{m}{2\pi}$  is the density of states of the two dimensional electrons at the Fermi level.

In addition, we want to consider dilute disorder which couples to the electromagnetic charge but not to the spin sector. A consistent treatment of such a system was presented in the context of the two dimensional disordered electron gas in a perpendicular magnetic field by Ando~\cite{Ando}, who incorporated this in the framework of the self-consistent Born approximation. On a phenomenological level in the oscillatory regime this can be done by introducing a finite lifetime for the electronic degrees of freedom. The one particle retarded Green's function reads
\begin{eqnarray}
G_f (\omega,n)&=&\frac{1}{\omega-\omega_{c}\left(n+\frac{1}{2}\right)+\frac{i}{2\tau}}\;.
\end{eqnarray}
The major effect of the disorder is thus to broaden the Landau levels into a Lorentzian shape. The regime of magneto-oscillations, which is what we are interested in, is characterized by $\omega_{c} \tau\ll1$. In this regime the density of states, for instance, has a constant part with a smooth oscillatory part on top of it, see Eq.~\eqref{eq:DOS}.

\subsection{Effective action for the gauge field in the diffusive limit}\label{polarization}
In order to calculate the specific heat in a $1/N$-expansion, we need to derive the photon propagator of the internal $U(1)$ gauge field in the presence of disorder and Landau levels. Schematically, our derivation goes along the following line: the fermionic action reads 
\begin{eqnarray}
\mathcal{S}= \int_0^\beta d \tau d^2x \overline{f} \left [i \partial_\tau + \mu -\epsilon (-i\nabla-e {\bf a} - {\bf A}) + i A_\tau \right] f \;.\nonumber \\
\end{eqnarray}
We assume a parabolic dispersion of the fermions, {\it i. e.} $\epsilon \left ({\bf k}\right)=\frac{{\bf k}^2}{2m}$. In a next step we integrate out the fermions, leading to 
\begin{eqnarray}
\mathcal{S}_{\rm eff}=-{\rm {tr}} \; {\rm {ln}} \; \left [i \partial_\tau +\mu  -\epsilon (-i\nabla-e {\bf a} - {\bf A}) + i a_\tau \right]\;. \nonumber \\
\end{eqnarray}
Expanding the above expression to second order in {\bf A} using $-i\nabla-e {\bf a}={\bf{\Pi}}$ we obtain the effective gauge field propagator from the polarization operator. In principle, one could derive the effective polarization operator in the presence of disorder in the basis of the Landau levels~\cite{Sakhi}. However, here we choose a different route.

In general, the polarization operator is obtained by expanding to second order in the internal gauge field and performing the functional derivative according to
\begin{eqnarray}
\frac{\delta^2 \mathcal{S}_{\rm eff}}{\delta {\bf A}^i ({\bf {q}},i\nu) \delta {\bf A}^j (-{\bf {q}},-i\nu) }={\hat{ \Pi}}^{ij}({\bf q},i\nu_n)\;.
\end{eqnarray}
Following Halperin {\it et al.}~\cite{Halperin} we keep the following two-component form of the photon-propagator
\begin{eqnarray}\label{eq:photonprop}
\hat{D}^{-1}({\bf q},\omega)=\hat{\Pi}{\bf q},\omega)=\left ( \begin{array}{cc}  {\hat{\Pi}}^{00} & \frac{q}{\omega} {\hat{\Pi}}^{xy} \\ -\frac{q}{\omega} {\hat{\Pi}}^{xy}  &  {\hat{\Pi}}^{yy} \end{array} \right) \;.
\end{eqnarray}
Subsequently, we concentrate on the low energy form of the photon propagator in the diffusive regime. The form of the photon propagator is highly constrained by conservation laws and generically reads
\begin{eqnarray}\label{eq:propelements}
\hat{\Pi}^{00}&=&\nu \frac{D q^2}{D q^2-i\omega}\;,\nonumber \\ \hat{\Pi}^{yy}&=&i\omega \sigma^{yy}+q^2 \chi \;,\nonumber \\ \hat{\Pi}^{xy}&=&i\omega \sigma^{xy}\;.
\end{eqnarray}
The diffusive limit of the system is accounted for by replacing $\Pi^{00}=\nu_0\to \nu_0 \frac{Dq^2}{Dq^2-i\omega}$ which on a formal level is achieved including impurity ladders in the vertex function. The dependence upon the magnetic field enters through $\sigma^{xx}$, $\sigma^{xy}$, $\nu$, and $\chi$. $D$ denotes the diffusion constant and is set by $D=\frac{v_{F}^2}{2}\tau$ in two dimensions. The free energy due to the gauge field is readily calculated (see Halperin {\it et al.}~\cite{Halperin})
\begin{eqnarray}\label{eq:gaugefree}
f^A&=&\int^{\Lambda} \frac{d^2 q}{(2\pi)^2} \int \frac{d\omega}{2\pi} n_b(\omega) \arctan  \frac{{\rm Im\;det}\; \hat{D}^{-1} ({\bf q},\omega)}{{\rm Re \;det}\; \hat{D}^{-1} ({\bf q},\omega)}\;.\nonumber \\
\end{eqnarray}
The upper cutoff $\Lambda$ is set by roughly twice the Fermi momentum, thus $\Lambda \approx 2 k_F$, which is the upper bound for the existence of low-energy excitations. 
Eq.~\eqref{eq:gaugefree} is the central expression which will allow to calculate the contribution of the internal gauge field with and without applied external magnetic field.

\section{Extracting the oscillatory part of the specific heat of the gauge field}\label{oscillatorygauge}

In the following we decompose the inverse photon propagator (Eq.~\eqref{eq:photonprop}) into two parts, one containing the non-oscillatory contributions, called $\hat{D}_0$ and another part containing the oscillatory contributions, called $\hat{D}_{\rm{osc}}$. The quantities $\sigma_{xx}$, $\sigma_{xy}$, $\chi$, $D$, and $\nu$ naturally decompose into a non-oscillatory part and an oscillatory part
\begin{eqnarray}
\nu&=&\nu_0+\nu^{\rm{osc}}\nonumber \\
\sigma_{xx}&=&\sigma^0_{xx}+\sigma^{\rm{osc}}_{xx}\nonumber \\
\sigma_{xy}&=&-\omega_c \tau \left ( \sigma^0_{xx}+\sigma^{\rm{osc}}_{xx}  \right )\nonumber \\ \chi&=&\chi^0+\chi^{\rm{osc}}
\end{eqnarray}
where all oscillatory contributions are expressed as a power series in $\exp \left(-\frac{\pi}{\omega_c \tau} \right)$, which on a formal level is obtained through a Poisson summation duality. As long as $\omega_c \tau \ll1$ it suffices to retain the first moment in this power series to isolate the leading oscillatory contribution. The results for the oscillatory components are given in the subsequent section (Sec.~\ref{oscquant}). 

Using the property from standard perturbation theory
\begin{eqnarray}\label{decomposition}
{\rm{det}}\; \hat{D}^{-1}&=&{\rm{det}}\; \hat{D}_0^{-1} {\rm{det}}\;(1+\hat{D}_0\cdot \hat{D}^{-1}_{\rm{osc}})\nonumber \\ &\approx&{\rm{det}}\; \hat{D}_0^{-1} (1+ {\rm{tr}}\;\hat{D}_0\cdot \hat{D}^{-1}_{\rm{osc}})
\end{eqnarray}
we can formulate the contribution of the gauge field to the free energy as
\begin{eqnarray}\label{freegauge}
f^A&=&\int^{\Lambda} \frac{d^2 q}{(2\pi)^2} \int \frac{d\omega}{2\pi} n_b(\omega) \arctan  \frac{{\rm Im\;det}\; \hat{D}^{-1} ({\bf q},\omega)}{{\rm Re\;det}\; \hat{D}^{-1} ({\bf q},\omega)}\nonumber \\ &\approx&\int^{\Lambda} \frac{d^2 q}{(2\pi)^2} \int \frac{d\omega}{2\pi} n_b(\omega) \arctan  \frac{{\rm Im\;det}\; \hat{D}_0^{-1} ({\bf q},\omega)}{{\rm Re\;det} \; \hat{D}_0^{-1} ({\bf q},\omega)}\nonumber \\ &+&\int^{\Lambda} \frac{d^2 q}{(2\pi)^2} \int \frac{d\omega}{2\pi} n_b(\omega) {\rm{Im}}\; {\rm{tr}}\; \hat{D}_0\cdot \hat{D}^{-1}_{\rm{osc}} \nonumber \\ &+&\mathcal{O}\left(e^{-\frac{2\pi}{\omega_c \tau}}\right)\;.
\end{eqnarray}
It is important to note that the second term is now proportional to $\exp \left({-\frac{\pi}{\omega_c\tau}}\right)$ and thus small compared to the first term. The first term in the above expression has been analyzed by Halperin et al.~\cite{Halperin}, and is known to yield a contribution to the free energy of the type $T \ln T$, which is reviewed later. The additional factor of $\ln (1/T)$ is the manifestation of the well-known Altshuler-Aronov~\cite{Altshuler} correction.

In the following we retain the leading order temperature dependence of the density-density and current-current response in $\mu/T$, however we will allow for arbitrary $T/\omega_c$.

\subsection{Oscillatory thermodynamic and transport input quantities}\label{oscquant}
The DOS can be calculated in a way analogous to the grand potential (see Appendix~\ref{App:grandpotential}). At the Fermi level, in the regime $\tau \mu \gg1$, the expression for the DOS reads
\begin{eqnarray}
\nu(\mu)=\nu_0 \left (1+2 \sum_{l=1}^\infty (-1)^l \cos \frac{2 \pi l \mu}{\omega_c} e^{-\frac{\pi l }{\omega_c \tau}} \right)\;,
\end{eqnarray}
whose leading oscillatory behavior in the limit $\omega_c \tau \ll 1$ reads
\begin{eqnarray}\label{eq:DOS}
\nu(\mu) \approx \nu_0 \left(1-2 \cos \frac{2 \pi \mu}{\omega_c} e^{-\frac{\pi}{\omega_c \tau}}\right)\;.
\end{eqnarray}
We can calculate the longitudinal conductivity accordingly~\cite{Ando}, yielding
\begin{eqnarray}
\sigma^{yy}&=&\sigma_0 \frac{1}{1+(\omega_c \tau)^2} \left (1+2 \sum_{l=1}^\infty (-1)^l \cos \frac{2 \pi l \mu}{\omega_c}  \frac{e^{-\frac{\pi l }{\omega_c \tau}}\lambda_l}{\sinh \lambda_l}\right) \nonumber \\ &\approx&\sigma_0 \frac{1}{1+(\omega_c \tau)^2} \left(1-2 \cos \frac{2 \pi \mu}{\omega_c} e^{-\frac{\pi}{\omega_c \tau}} \frac{\lambda_1}{\sinh \lambda_1}\right)\;,\nonumber \\
\end{eqnarray}
with $\sigma_0=\frac{ne^2 \tau}{m}$ and $\lambda_l=\frac{2 \pi^2 T l}{\omega_c}$. 

Using the well known Einstein relation for diffusive systems
\begin{eqnarray}
\sigma=\nu D
\end{eqnarray}
we can determine the oscillatory part of the diffusion constant $D$. It turns out that to leading order we have
\begin{eqnarray}
D=D_0\left(1+2 \cos \frac{2\pi \mu}{\omega_c}e^{-\frac{\pi}{\omega_c \tau}}\left(1- \zeta(T,\omega_c)\right)\right)\;,
\end{eqnarray}
where we introduced the function
\begin{eqnarray}
\zeta(T,\omega_c)=\frac{2 \pi^2 T/\omega_c}{\sinh 2 \pi^2 T/\omega_c}
\end{eqnarray}
for notational convenience.
We observe that for $T \ll \omega_c$ $D=D_0$ to all orders in ${\rm{exp}}\left(-\frac{\pi}{\omega_c \tau}\right)$. Furthermore, it is straightforward to show that
\begin{eqnarray}
\sigma_{xy}=-\omega_c \tau \sigma_{xx}\;,
\end{eqnarray}
to all orders in $\exp \left(-\frac{\pi}{\omega_c \tau}\right)$.

The diamagnetic susceptibility can be obtained from the grand potential (Eq.~\eqref{eq:freeenergy}) by
\begin{eqnarray}
\chi=-\frac{\partial^2 \Omega}{\partial B^2}\;.
\end{eqnarray}
Using Eq.~\eqref{eq:freeenergy} in the limit $\omega_c \tau \ll 1$, $\frac{\omega_c}{\mu}\ll1$, and $\mu \tau \gg 1$ we obtain
\begin{eqnarray}
\chi &\approx&- \frac{1}{24 \pi m}\left(1+24 \frac{\mu^2}{\omega_c^2} \cos \frac{2\pi \mu}{\omega_c} e^{-\frac{\pi}{\omega_c \tau}} \zeta(T,\omega_c) \right) \nonumber \\ &=& \chi_0 \left(1+ 24 \frac{\mu^2}{\omega_c^2} \cos \frac{2\pi \mu}{\omega_c} e^{-\frac{\pi}{\omega_c \tau}}\zeta(T,\omega_c)   \right)\;.
\end{eqnarray}

 \section{Free energy and specific heat}\label{freespec}
In the spirit of the large-$N$ treatment we can expand the free energy $\mathcal{F}$ of the system to order $1/N$, which yields the following result
\begin{eqnarray}
\mathcal{F}=N f^{0f}+f^{A} \;,
\end{eqnarray}
where $f^{0f}$ is the free energy of the non-interacting fermionic system and $f^A$ denotes the free energy associated with the fluctuations of the emergent gauge field.

The specific heat of this expression can be obtained by the well-known formula $c_V=-T\frac{\partial^2 \mathcal{F}}{\partial T^2}$. This implies that the specific heat decomposes into two parts
\begin{eqnarray}
c_V=N c_V^{f}+c_V^{A}\;,
\end{eqnarray}
which are analyzed independently.

\subsection{Specific heat of the electrons}\label{specelec}
The free energy is related to the grand potential via Legendre transform according to
\begin{eqnarray}
f^{0f}=\mu N + \Omega \;,
\end{eqnarray}
which allows to obtain the specific heat via 
\begin{eqnarray}
c_V=-T \frac{\partial^2 f^{0f}}{\partial T^2}\;.
\end{eqnarray}
The dependence of the chemical potential upon the magnetic field is subdominant, thus 
\begin{eqnarray}
c_V=-T\frac{\partial^2 \Omega}{\partial T^2}
\end{eqnarray}
with the oscillatory contribution obtained from Eq.~\eqref{eq:freeenergy}
\begin{eqnarray}\label{fermosc}
c_V^{\rm{osc}}&=& \alpha(T) \frac{\nu_0}{2\pi^2} T \cos \frac{2\pi \mu}{\omega_c} e^{-\frac{ \pi }{\omega_c \tau}}\nonumber \\ &=& \alpha(T) \frac{m T}{4\pi^3}\cos \frac{2\pi \mu}{\omega_c} e^{-\frac{ \pi }{\omega_c \tau}}\;,
\end{eqnarray}
where
\begin{eqnarray}
\alpha(T) &=&\omega_c^2  \frac{\partial^2}{\partial T^2} f(T,\omega_c)\nonumber \\
&=&  \frac{4 \pi^6 T }{\omega_c \sinh^3 \left( \frac{2 \pi^2 T}{\omega_c} \right)}\left ( 3+\cosh \left (\frac{4\pi^2 T}{\omega_c} \right) \right) \nonumber \\ &-&\frac{4 \pi ^4  \sinh \left(\frac{4 \pi ^2T}{\omega_c}\right)}{\sinh^3\left(\frac{2 \pi ^2T}{\omega_c}\right)} \;.
\end{eqnarray}
This function is dimensionless and compared to the oscillatory contribution from the gauge field in Fig.~\ref{fig1}.

In the limit $\frac{T}{\omega_c}\ll1$ this reduces to the well-known 
\begin{eqnarray}\label{cvelectrons}
c_V=\frac{\pi^2 \nu_0}{3}T \left(1-2 \cos \frac{2\pi \mu}{\omega_c}e^{-\frac{\pi}{\omega_c \tau}} \right)
\end{eqnarray}
per spin species. 
\subsection{Specific heat of the gauge field}\label{specgauge}

In the following we calculate the non-oscillatory and oscillatory contributions of the gauge field to the specific heat separately. 

\subsubsection{Non-oscillatory contribution}\label{nonosc}

We start with a calculation of the non-oscillatory part of the specific heat of the gauge field. In order to do so we analyze
\begin{eqnarray}
\int^{\Lambda} \frac{d^2 q}{(2\pi)^2} \int \frac{d\omega}{2\pi} n_b(\omega) \arctan  \frac{{\rm det}\; \hat{D}_0^{-1} ({\bf q},\omega)''}{{\rm det} \; \hat{D}_0^{-1} ({\bf q},\omega)'}\;.
\end{eqnarray}
We can rewrite the contribution to the specific heat as
\begin{eqnarray}
c_V^{A}&=&\frac{T}{16\pi^2} \int_0^{\Lambda/\sqrt{T}} dk \int_{-\infty}^{\infty}  d x   \nonumber \\ &&\frac{k\left(2x-x^2 \coth \left(\frac{x}{2}\right)\right)}{\sinh(x/2)^2}\arctan\frac{cxk^2}{ak^4+bx^2}\;,
\end{eqnarray}
with
\begin{eqnarray}
a&=& D\chi_0 (1+48 \pi^2 (\omega_c \tau)^2\left(\sigma_0^{yy}\right)^2)\;, \nonumber \\ b&=& -\sigma^{yy}_0(1+\omega_c^2 \tau^2)\approx -\sigma^{yy}_0\;, \nonumber \\ c&=& D \sigma^{yy}_0+\chi_0 \approx D \sigma^{yy}_0  \;.
\end{eqnarray}
for $\omega_c \tau\ll1$ and $\mu \tau \gg1$.
In order to analyze the asymptotic behavior of the above $x$-integral, we start noting that the $x$-integration is cut off by the factor $\frac{1}{\sinh^2 \frac{x}{2}}$ on the order of $x=10$. If we consider the factor $\frac{1}{a k^4+b x^2}$, we know that for $k \gg \sqrt{10}\left(\frac{b}{a}\right)^{1/4}$ it becomes $\frac{1}{a k^4}$, whereas for $k \ll 4$ it becomes $\frac{1}{x^2}$. Only in the former case will the integral contribute a logarithmic dependence upon temperature, hence
\begin{eqnarray}
c_V^{A}\approx-\frac{cT}{6a}\int_{\sqrt{10}\left(\frac{b}{a}\right)^{1/4}}^{\Lambda/\sqrt{T} }\frac{dk}{k}=-\frac{cT}{12a}\ln \frac{\Lambda^2 a^{1/2}}{10 b^{1/2}T}\;.
\end{eqnarray}
 We furthermore introduce the short form 
 \begin{eqnarray}\label{kappa}
 \kappa=1+48 (\mu \tau)^2 (\omega_c \tau)^2\;,
\end{eqnarray}
which is a number of order $1$.
This leaves us with
\begin{eqnarray}\label{gaugenonosc}
c_V^A&=&8 \frac{\mu \tau m}{\kappa} T \ln \left( \frac{8 \mu}{5 T}\sqrt{\frac{\kappa}{3}} \right)
\end{eqnarray}
as the non-oscillatory contribution.

\subsubsection{Oscillatory contribution of the gauge field}\label{oneelec}
In this section we analyze the oscillatory contribution of the gauge field. 

\begin{figure}
\includegraphics[width=0.45\textwidth]{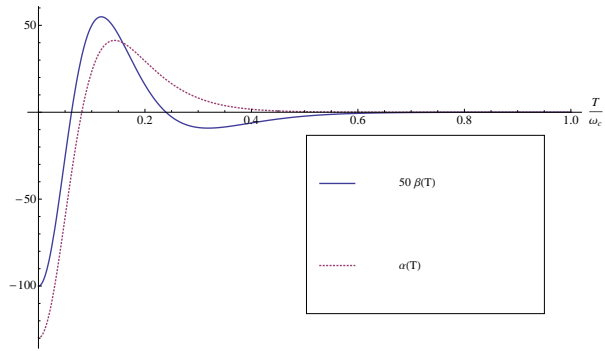}
\caption{ This plot shows the function $\alpha(T)$ and $50 \beta(T)$, where T is measured in units of $\omega_c$ and $\frac{4\kappa}{3\tau} =100$ in units of $\omega_c$. We magnified the function $\beta(T)$ by the arbitrary factor $50$ for a better comparison of the two curves. We see that both curves saturate for very low temperatures.}\label{fig1}
\end{figure}

Following the prescription given in Eq.~\eqref{decomposition} we decompose the gauge-field propagator according to
\begin{eqnarray}
\hat{D}^{-1} &=&\left ( \begin{array}{cc} \nu \frac{D q^2}{D q^2-i\omega} & i q \sigma^{xy}\\ -i q\sigma^{xy} & i\omega \sigma^{yy}+\chi q^2 \end{array} \right) \nonumber \\ &=&
\left ( \begin{array}{cc} \nu_0 \frac{D_0 q^2}{D_0 q^2-i\omega} & i q \sigma^{xy}_0\\ -i q\sigma^{xy}_0 & i\omega \sigma^{yy}_0+\chi_0 q^2 \end{array} \right)\nonumber \\ &+&\left ( \begin{array}{cc} \left ( \nu \frac{D q^2}{D q^2-i\omega}\right)_{\rm{osc}} & i q \sigma^{xy}_{\rm{osc}}\\ -i q\sigma^{xy}_{\rm{osc}} & i\omega \sigma^{yy}_{\rm{osc}}+\chi_{\rm{osc}} q^2 \end{array} \right)\nonumber \\ &=&\hat{D}_0^{-1}+\hat{D}_{\rm{osc}}^{-1}\;.
\end{eqnarray}
Using the oscillatory expressions derived in Section~\ref{oscquant} we identify
\begin{eqnarray}\label{Dosc}
&&\hat{D}_{\rm{osc}}^{-1} \approx -2\cos\frac{2\pi\mu}{\omega_c}e^{-\frac{\pi}{\omega_c \tau}} \hat{D}_0^{-1} +\cos\frac{2\pi\mu}{\omega_c}e^{-\frac{\pi}{\omega_c \tau}}\times \nonumber \\ &\times& \left ( \begin{array}{cc} -2 \frac{\nu_0 D_0^2 q^4}{(D_0 q^2-i \omega)^2} \left( 1- \zeta(T,\omega_c)\right) & 0 \\ 0 &24 \frac{\mu^2}{\omega_c^2} \chi_0 q^2 \zeta(T,\omega_c)  \end{array}\right), \nonumber \\ 
\end{eqnarray}
where $\frac{\omega_c}{\mu}\ll1$ was used.
Calculating ${\rm{Im}}\;{\rm{tr}}\; \hat{D}_0\cdot \hat{D}_{\rm{osc}}^{-1}$ we realize that the first term on the right hand side of Eq.~\eqref{Dosc} does not contribute an imaginary part, thus 
\begin{eqnarray}\label{eq:traceexpand}
&&{\rm{Im}}\;{\rm{tr}}\; \hat{D}_0\cdot \hat{D}_{\rm{osc}}^{-1}=- 24 \frac{\mu^2}{\omega_c^2} \cos\frac{2\pi\mu}{\omega_c}e^{-\frac{\pi}{\omega_c \tau}}\times \nonumber \\ &\times& \left ( f(T,\omega_c) \frac{C q^2 \omega}{A q^4+B \omega^2} \right.  \nonumber \\  &+& \left. \frac{\omega_c^2}{12\mu^2}\left ( f(T,\omega_c)-1\right) \frac{\tilde{C} q^2 \omega}{A q^4+B \omega^2} \frac{\tilde{a}q^4+\tilde{b}\omega^2}{D^2 q^4+\omega^2}\right) .\nonumber \\
\end{eqnarray}
Looking at Eq.~\eqref{eq:traceexpand} it is interesting to note that there are two contributions. The first contribution survives in the limit $T/\omega_c \to 0$, whereas the second part goes to zero. Additionally, the second term is down by a factor $\frac{\omega_c^2}{\mu^2}$ and is thus parametrically small compared to the first term. Consequently, we will discard the second contribution for our analysis. The remaining constants read
\begin{eqnarray}
A&=& D^2 \left(\left( \sigma^{xy}_0\right)^2-\nu_0 \chi_0  \right)^2 \;, \nonumber \\ B&=&  \left(\left( \sigma^{xy}_0\right)^2+\left( \sigma^{yy}_0\right)^2  \right)^2 \approx \left( \sigma^{yy}_0\right)^4   \;, \nonumber \\ C&=& D \nu_0 \chi_0 \left(\left( \sigma^{xy}_0\right)^2+\left( \sigma^{yy}_0\right)^2  \right)   \approx D \nu_0 \chi_0 \left( \sigma^{yy}_0\right)^2  \;.\nonumber \\
\end{eqnarray}
The constants of the disordered electronic system involved in the above expression assume the well-known values
\begin{eqnarray}
\sigma_{0}^{yy}&=&\frac{ne^2}{m}\tau \frac{1}{1+(\omega_c \tau)^2}=\frac{1}{1+(\omega_c \tau)^2} \sigma_0 \approx \sigma_0\;, \nonumber \\ \sigma_{0}^{xy}&=&-\frac{\omega_c \tau}{1+(\omega_c \tau)^2} \sigma_0  \approx -\omega_c \tau \sigma_0\;, \nonumber \\ \nu_0&=& \frac{m}{2\pi}\;, \quad \chi_0 =-\frac{1}{24 \pi m}\;, \nonumber \\ \frac{\mu^2}{\omega_c^2}&=&\left( \frac{2 \pi}{ \omega_c \tau}\right)^2 \sigma_0^2\;, \quad D=\frac{2 \pi \sigma_0}{m}\;.
\end{eqnarray}
We now calculate the temperature dependent part of the free energy (see Appendix~\ref{app:photon}). This part reads
\begin{eqnarray}
f^{A}_{\rm{osc}}&=&-  \frac{\mu^2}{\omega_c^2} \frac{C}{A} \cos\frac{2\pi\mu}{\omega_c}e^{-\frac{\pi}{\omega_c \tau}} \zeta(T,\omega_c)T^2 \ln \left ( \sqrt{\frac{A}{B}}\frac{\Lambda^2}{T}\right) \nonumber \\ &-&\xi \frac{\mu^2}{\omega_c^2}\frac{C}{A}\cos\frac{2\pi\mu}{\omega_c}e^{-\frac{\pi}{\omega_c \tau}} \zeta(T,\omega_c) T^2\;, \nonumber \\
\end{eqnarray}
with $\xi\approx0.352822$.
We introduce the dimensionless functions
\begin{eqnarray}
\beta(T)&=&- \frac{\partial^2}{\partial T^2} T^2 \zeta(T,\omega_c)\;, \nonumber \\
\gamma(T)&=& - \zeta (T,\omega_c)+\frac{1}{T}\frac{\partial }{\partial T} T^2 \zeta (T,\omega_c)+\xi \beta(T)\;,
\end{eqnarray}
which allows to express the specific heat as
\begin{eqnarray}\label{gaugeosc}
c_V^{A\rm{osc}} &=& \cos\frac{2\pi\mu}{\omega_c} e^{-\frac{\pi}{\omega_c \tau}} \frac{24 \left(\mu \tau\right)^3 }{ \left(\omega_c \tau\right)^2\kappa^2}m T  \ln \left ( \sqrt{\frac{A}{B}}\frac{\Lambda^2}{T}\right)  \beta(T)   \nonumber \\ &+& \cos\frac{2\pi\mu}{\omega_c} e^{-\frac{\pi}{\omega_c \tau}} \frac{24 \left(\mu \tau\right)^3 }{ \left(\omega_c \tau\right)^2\kappa^2}m T \gamma(T)\;, 
\end{eqnarray}
with $\kappa$ as defined in Eq.~\eqref{kappa}. If we compare the oscillatory contribution of the LK type (Eq.~\eqref{fermosc}) with the $1/N$ corrections we realize that their ratio behaves as
\begin{eqnarray}
\frac{c_V^{A\rm{osc}}(T)}{c_V^{\rm{osc}}(T)}=\frac{96 \pi^3 (\mu \tau)^3}{N(\omega_c \tau)^2 \kappa^2} \frac{\beta(T)\ln \left ( \sqrt{\frac{A}{B}}\frac{\Lambda^2}{T}\right)+\gamma(T)}{\alpha(T)}\;.
\end{eqnarray}
For low temperatures the logarithmically diverging part dominates the $1/N$ correction, which is why we plot the functions $\alpha(T)$ and $\beta (T)$ only. 
For a reasonable qualitative comparison of the functional forms of the two contributions we take the dimensionless prefactor to be of the order $50$. Assuming this number, Fig.~\ref{fig1} shows a comparison of the temperature dependent dimensionless functions $\alpha(T)$ and $\beta(T)$. However, one still has to keep in mind that the $1/N$-correction has an additional logarithmic temperature dependence, which was scaled out for the comparison.
\section{Application to the underdoped cuprate superconductors}\label{sec:cuprates}

Recent experiments performed in the underdoped regime of the cuprate superconductors show great promise to shed light onto these still mysterious materials. In our discussion we focus on quantum oscillations measurements in the underdoped region of YBa$_2$Cu$_3$O$_{7-\delta}$ (YBCO)~\cite{Doiron,Yelland,Bangura,Jaudet,Sebastian,LeBoeuf}. In this context, LeBoeuf {\it et al.}~\cite{LeBoeuf} observed signatures reminiscent of pockets of carriers of charge $-e$ (in contrast to hole-like charge-carriers). 

Here, we will discuss these experiments using a specific theoretical model~\cite{Kaul,Galitski,Moon,SachdevDresden,Xu} for the interplay between spin density wave (SDW) and superconductivity
in the underdoped regime. We investigate the magneto oscillations for magnetic fields 
greater than $H_{c2}$, {\it i.e.} in the absence of superconductivity. 
In contrast to earlier works~\cite{Galitski} which investigated Shubnikov-de-Haas oscillations, we emphasis the role of de-Haas-van-Alphen oscillations, {\it i.e.} oscillation in thermodynamic quantities.

The theory has two non-superconducting metallic phases. One has long-range SDW order, and so is a conventional
Fermi liquid at low enough temperatures: here the magneto oscillations will be given by the LK theory.
The other metallic phase has no SDW order, but retains aspects of the Fermi pocket structure of the SDW-ordered
phase: this is the `algebraic charge liquid'. The ACL has an emergent gapless U(1) photon which will lead
to corrections to the LK theory, as we have discussed above. The photon acquires a Higgs mass, $\Delta_{AF}$, across
the transition from the ACL to the SDW phase, and so its effects are quenched in the SDW phase.

The specific model has the Lagrangian
\begin{eqnarray}
\mathcal{L}=\mathcal{L}_z+\mathcal{L}_g+\mathcal{L}_f\;.
\end{eqnarray}
The first term describes the magnetic degrees of freedom. Conventionally, the slow magnetic degrees of freedom are expressed in the framework of $O(3)$ non-linear sigma model. Here, however, we map the spin density wave (SDW) order parameter $\vec{\phi}$ to bosonic degrees of freedom $z$, which carry spin $S=\frac{1}{2}$, via
\begin{eqnarray}\label{eq:Hopf}
\vec{\phi}=\sum_{\alpha,\beta}z^\star_\alpha \vec{\sigma}_{\alpha \beta} z^{\phantom{\star}}_{\beta}
\end{eqnarray}
and the effective Lagrangian assumes the form of the so-called $CP^1$-model
\begin{eqnarray}
\mathcal{L}_z&=&\frac{1}{t}\left [\sum_{\alpha=1}^N \left( |(\partial_\tau-iA_\tau)z_{\alpha}|^2+v^2|({\bf{\nabla}}-i{\bf{A}})z_\alpha|^2 \right) \right. \nonumber \\ &+& \left. i\rho \sum_{\alpha=1}^N \left(|z_\alpha|^2-N  \right)  \right]\;,
\end{eqnarray}
where $A_\tau$ and ${\bf{A}}$ denote an internal U(1) gauge field emerging from the redundant parametrization of the SDW order parameter shown in Eq.~\eqref{eq:Hopf}. Interestingly, the bosonic spinons $z$ locally determine the spin axis of the physical electrons, which leads to a fractionalization of the spin and the charge degree of freedom. Consequently, the effective charge carriers also couple to the internal gauge field. It was known for a long time that the existence of spin density wave order, {\it i.e.} $\langle z\rangle \neq 0$ is responsible for a Fermi surface reconstruction~\cite{Sokol,Chubukov}. In a very simple approximation we can take this fact into account by introducing $g_+$ and $g_-$ for the electronic pockets sitting at the antinodal points (($\pi,0$))and $f_{+1}$, $f_{+2}$, $f_{-1}$, and $f_{-2}$ for the hole degrees of freedom sitting at the nodal points (($\pi/2,\pi/2$)):
\begin{eqnarray}\label{eq:ferm1}
\mathcal{L}_g &=& \overline{g}_+ \left( \partial_\tau-iA_\tau-\frac{(\nabla-i{\bf{A}})^2}{2m_1}-\mu\right)g_+ \nonumber \\ &+& \overline{g}_- \left( \partial_\tau+iA_\tau-\frac{(\nabla+i{\bf{A}})^2}{2m_1}-\mu\right)g_- \nonumber \\
\end{eqnarray}
and
\begin{eqnarray}\label{eq:ferm2}
\mathcal{L}_f &=& \sum_{q,a}\overline{f}_{qa} \left( \partial_\tau-iqA_\tau-\frac{(\nabla-iq{\bf{A}})^2}{2m_2}-\mu\right)f_{qa}\;,\nonumber \\
\end{eqnarray}
where $q=\pm$ and $a=1,2$.
As explained before, all the fermionic degrees are coupled to the internal $U(1)$ gauge field, but carry different charges under the transformation.

The generic phase diagram of the above model has been worked out in very recent publications~\cite{Moon,SachdevDresden} and is not repeated here. Our work concentrates on the metallic SDW and ACL states with small Fermi pockets, {\it i.e.} for magnetic fields $H$ bigger than the critical field strength $H_{c2}$ above which superconductivity is destroyed.

In the above model we again introduce disorder and magnetic field on the level of the single particle propagator. We realize that there are two cyclotron frequencies $\omega_{cg}$ and $\omega_{cf}$, associated with the two different sorts of electrons. Additionally, there are two scattering times $\tau_g$ and $\tau_f$. The following discussion is again valid in the limit $\omega_{cg}\tau_g,\omega_{cf}\tau_f \ll1$. The specific heat of the system is this time composed of four different contributions
\begin{eqnarray}
c_V=N_z c_V^z+N_g c_V^g+N c_V^f+c_V^A
\end{eqnarray}
The first term is due to the bosonic spinons and was calculated in Ref.~\cite{RKKaul}. The two following terms were calculated together with the magneto-oscillations in Eq.~\eqref{cvelectrons}. Turning to the gauge field propagator we realize that the gauge field propagator is still given by Eq.~\eqref{eq:photonprop}, however Eq.~\eqref{eq:propelements} modifies to
\begin{eqnarray}\label{eq:inputfull}
\hat{\Pi}^{00}&=&\nu_g \frac{D_g q^2}{D_g q^2-i\omega}+\nu_f \frac{D_f q^2}{D_f q^2-i\omega}\;,\nonumber \\ \hat{\Pi}^{yy}&=&i\omega \sigma^{yy}_g+q^2 \chi_g + i\omega \sigma^{yy}_f+q^2 \chi_f\;,\nonumber \\ \hat{\Pi}^{xy}&=&i\omega \sigma^{xy}_g+i\omega \sigma^{xy}_f\;,
\end{eqnarray}
which implies we can write
\begin{eqnarray}
\hat{D}^{-1}=\hat{D}_g^{-1}+\hat{D}_f^{-1}\;.
\end{eqnarray}
A further modification comes into the picture due to the presence of the bosonic spinons. In the SDW state they condense, {\it i.e.} $\langle z \rangle^2 \sim \Delta_{AF}$, implying that the Higgs effect contributes a mass term causing $\hat{\Pi}^{yy}=i\omega \sigma^{yy}_g+q^2 \chi_g + i\omega \sigma^{yy}_f+q^2 \chi_f+\Delta_{AF}$. 

For the moment, we will neglect this term and defer the discussion of the SDW state to the end of the section. As we discussed earlier, the oscillatory contributions to the thermodynamic and transport quantities entering Eq.~\eqref{eq:inputfull} are of the form $e^{-\frac{\pi}{\omega_{cg}\tau_g}}$ and $e^{-\frac{\pi}{\omega_{cf}\tau_f}}$. In the following we will consider $e^{-\frac{\pi}{\omega_{cf}\tau_f}} \ll e^{-\frac{\pi}{\omega_{cg}\tau_g}}$. With this we can express the oscillating term to leading order in $\frac{\mu}{\omega_c}$ according to
\begin{eqnarray}
\hat{D}_{\rm{osc}}^{-1}&=&-2\cos\frac{2\pi\mu}{\omega_{cg}}e^{-\frac{\pi}{\omega_{cg} \tau_g}} \hat{D}_0^{-1} \nonumber \\ &+&24 \frac{\mu^2}{\omega_{cg}^2}\cos\frac{2\pi\mu}{\omega_{cg}}e^{-\frac{\pi}{\omega_{cg} \tau_{g}}} \zeta(T,\omega_{cg}) \left ( \begin{array}{cc} 0 & 0 \\ 0 & \chi_g  q^2  \end{array}\right) \nonumber \\ &+&\mathcal{O}\left (e^{-\frac{2\pi}{\omega_{cg}\tau_g}},e^{-\frac{\pi}{\omega_{cf}\tau_f}},\left ( \frac{\mu}{\omega_c}\right)^0 \right)\;.\nonumber \\
\end{eqnarray}
As in our preceeding discussion in Sec.~\ref{oneelec}, the first term drops out and eventually we can write the whole expression as
\begin{eqnarray}
&&{\rm{Im}}\; {\rm{tr}}\;\hat{D}_0 \cdot \hat{D}_{\rm{osc}}^{-1}=\nonumber \\ &&- 24\frac{C q^6 \omega+\tilde{C} q^2 \omega^3}{A q^8+\tilde{A} q^4 \omega^2+B \omega^4}\frac{\mu^2}{\omega_{cg}^2}\cos \frac{2\pi \mu}{\omega_{cg}}e^{-\frac{\pi}{\omega_{cg}\tau_g}} \nonumber \\ 
\end{eqnarray}
Again, we are only interested in isolating the logarithmic behavior, for which we only need to know the constants $A$, $B$, and $C$. We find following the earlier calculation that the logarithmic part of the specific heat is given by
\begin{eqnarray}\label{cvhightc}
c_V^{A\rm{osc}}= &-&\frac{C}{A} \frac{\mu^2}{\omega_{cg}^2} \cos \frac{2\pi \mu}{\omega_{cg}}e^{-\frac{\pi}{\omega_{cg}\tau_g}} \times \nonumber \\ &\times&\ln \left(\sqrt{\frac{A}{B}} \frac{\Lambda^2}{T}\right) T \frac{\partial^2}{\partial T^2}T^2 \zeta(T,\omega_{cg})\nonumber \\
\end{eqnarray}
with
\begin{eqnarray}
A&=& D_g^2 D_f^2 \left(\left(\sigma^{xy}\right)^2-\nu \chi\right)^2\;,  \nonumber \\ B&=& \left(\sigma^{yy}\right)^4 \;, \nonumber \\ C&=&D_g^2 D_f^2   \nu^2 \sigma^{yy} \chi_g \;,
\end{eqnarray}
where 
\begin{eqnarray}
\sigma^{xy} &=& \sigma^{xy}_g+ \sigma^{xy}_f \;, \nonumber \\ \sigma^{yy} &=& \sigma^{yy}_g+ \sigma^{yy}_f\;,\nonumber \\ \nu &=& \nu_g+\nu_f\;, \nonumber \\ \chi&=& \chi_g+\chi_f\;.
\end{eqnarray}
In the SDW phase, the presence of a finite Higgs term $\Delta_{AF}\neq0$ in the gauge field propagator introduces a new energy scale into the problem and the character of the problem changes its character to that of a Fermi liquid and the magneto oscillations will essentially be given by LK theory. Without going into the details we can show this along the lines of the derivation of Eq.~\eqref{cvhightc}. The logarithmically in temperature diverging prefactor is now cut off by the Higgs mass $\Delta_{AF}$.  A compact formulation of Eq.~\eqref{cvhightc} treating both regimes is given by
\begin{eqnarray}\label{cvhightcsdw}
c_V^{A\rm{osc}}= &-&\frac{C}{A} \frac{\mu^2}{\omega_{cg}^2} \cos \frac{2\pi \mu}{\omega_{cg}}e^{-\frac{\pi}{\omega_{cg}\tau_g}} \times \nonumber \\ &\times&\ln \left(\sqrt{\frac{A}{B}} \frac{\Lambda^2}{\rm{max}\left[\frac{\Delta_{AF}}{\chi},T\right]}\right) T \frac{\partial^2}{\partial T^2}T^2 f(T,\omega_{cg}) \;,\nonumber \\
\end{eqnarray}
which in the SDW phase at very low temperatures ($T\ll \frac{\Delta_{AF}}{\chi}$) corresponds to a simple renormalization of Fermi liquid theory.
\section{Summary and discussion}\label{summary}

Within this paper we analyzed magneto-oscillations in the specific heat of two-dimensional non-relativistic fermions coupled to a $U(1)$ gauge field. Our results apply to a variety of different problems, including the description of the $\nu=1/2$ fractional quantum Hall state~\cite{Halperin,Kalmeyer,Simon} and different effective low energy gauge theory descriptions of the cuprate superconductors~\cite{Ioffe,Nagaosa,Lee,Galitski}. Our main result obtained in this paper is the calculation of the correction to the standard Fermi liquid result of Eq.~\eqref{cvelectrons}, which is shown in Eq.~\eqref{gaugeosc}. We point out a couple of generic situations, in which our result holds and furthermore discuss its meaning in the context of the gauge theoretic description of the underdoped cuprates introduced by Galitski and Sachdev~\cite{Galitski}.
 
\acknowledgments We acknowledge useful discussions with A. Altland, G. S. Boebinger, S. Florens, B. Halperin, M. A. Levin and S. Riggs. This research was supported by the Deutsche Forschungsgemeinschaft under grant FR 2627/1-1 (LF), and by the NSF under grant DMR-0757145 (SS, LF).

\appendix
\section{The grand potential}\label{App:grandpotential}
The grand potential of the disordered Fermi gas in a magnetic field is readily calculated using the fermionic propagator
\begin{eqnarray}
\Omega=-\beta^{-1}\frac{1}{2 \pi l_B^2} \sum_{m,n}\ln \left ( -G^{-1}(\omega_n,E_m)\right)
\end{eqnarray}
where 
\begin{eqnarray}
G(\omega_n,E_m)=\frac{1}{i\omega_n+\mu-E_m+i {\rm{sgn}}\omega_n\frac{1}{2\tau}}
\end{eqnarray}
with $E_m=\omega_c(m+1/2)$.
We use the following Poisson summation identity~\cite{Shoenberg}
\begin{eqnarray}
&&\sum_{m} f\left (\omega_c \left (m+\frac{1}{2}\right)\right)=\int_0^{\infty} \frac{dx}{\omega_c} f(x)\nonumber \\ &-&2 \sum_{l=1}^\infty \frac{(-1)^l }{2\pi l} \int_0^\infty dx f'(x) \sin \frac{2 \pi l x}{\omega_c} 
\end{eqnarray}
with 
\begin{eqnarray}
f'(x)&=&\frac{-1}{i\omega_n+\mu-x+i\frac{{\rm{sgn}}\omega_n}{2\tau}}\nonumber \\ &=&-\frac{1}{2 \pi \tau} \int_{-\infty}^\infty d\omega \frac{1}{i\omega_n-\omega} \frac{1}{(\omega+\mu-x)^2+\frac{1}{4\tau^2}} \;. \nonumber \\
\end{eqnarray}
We can now perform the Matsubara sum yielding
\begin{eqnarray}
\beta^{-1} \sum_n f'(x)=\int_{-\infty}^\infty d \omega n_f(\omega) G''(\omega)\;.
\end{eqnarray}
We furthermore use $\frac{1}{2\pi l_B^2}=\nu_0 \omega_c$, which implies
\begin{eqnarray}
\Omega&=&-\beta^{-1}\sum_n \nu_0 \int_0^\infty dE \ln (-i\omega_n+\mu-E-i\frac{{\rm{sgn}}\omega_n}{2\tau})\nonumber \\ &+&2\nu_0 \omega_c \sum_{l=1}^\infty \frac{(-1)^l}{2\pi^2 l} \int_0^\infty dE \sin \frac{2\pi l E}{\omega_c}\times \nonumber \\ &\times& \int d\omega n_f(\omega) G''(\omega,E)\;. \nonumber \\
\end{eqnarray}
We will now concentrate on calculating the second term. We integrate the energy integration by parts to obtain
\begin{eqnarray}
&&\int_0^\infty dE \sin \frac{2\pi l E}{\omega_c} \int d\omega n_f(\omega) G''(\omega,E)\nonumber \\&=&\int d\omega n_f(\omega) \frac{\omega_c}{2\pi l} \cos \frac{2\pi l E}{\omega_c} (G''(\omega,0)-G''(\omega,\infty))  \nonumber \\ &-&\int d\omega n_f(\omega) \frac{\omega_c}{2\pi l} \int_0^\infty dE \cos \frac{2\pi l E}{\omega_c} \frac{d}{dE}G''(\omega,E)\;.\nonumber \\
\end{eqnarray}
The first term on the right-hand only produces one term. This can easily be checked, since for $E=0$ the expression is finite, whereas for $E\to \infty$ one can see, that the term vanishes, since the Fermi function restricts $\omega$ to be smaller than zero. Consequently, we obtain
\begin{eqnarray}
&&\int_0^\infty dE \sin \frac{2\pi l E}{\omega_c} \int d\omega n_f(\omega) G''(\omega,E)\nonumber \\&=&\int d\omega n_f(\omega) \frac{\omega_c}{2\pi l} G''(\omega,0)  \nonumber \\ &-&\int d\omega n_f(\omega) \frac{\omega_c}{2\pi l} \int_0^\infty dE \cos \frac{2\pi l E}{\omega_c} \frac{d}{dE}G''(\omega,E)\;.\nonumber \\
\end{eqnarray}
We can split the remaining task into two parts. For the first integral one obtains in the limit $\mu \gg \delta,T$
\begin{eqnarray}
\int d\omega n_f(\omega) \frac{\omega_c}{2\pi l}G''(\omega,0) =-\frac{\omega_c}{2 l}    \;.
\end{eqnarray}
We now calculate the second integral
\begin{eqnarray}
&-&\int d\omega n_f(\omega) \frac{\omega_c}{2\pi l} \int_0^\infty dE \cos \frac{2\pi l E}{\omega_c}\frac{d}{dE}G''(\omega,E)\nonumber \\ &=&  n_f(\omega) \frac{\omega_c}{2\pi l} \int_0^\infty dE \cos \frac{2\pi l E}{\omega_c} (G''(\omega,-\infty)-G''(\omega,\infty)) \nonumber \\ &+&\int d\omega \frac{d n_f(\omega)}{d \omega} \frac{\omega_c}{2\pi l} \int_0^\infty dE \cos \frac{2\pi l E}{\omega_c}G''(\omega,E) \;.
\end{eqnarray}
The first term again is easily analyzed. For $\omega\to \infty$ the Fermi distribution annihilates the expression. For $\omega \to -\infty$ we realize that the denominator of the Green's function diverges, since $E>0$, which implies that the denominator overall goes like $\frac{1}{(\omega+\mu)^2+\frac{1}{4\tau^2}}$ for $\omega \to -\infty$, thus going quadratically to zero. This implies
\begin{eqnarray}
&-&\int d\omega n_f(\omega) \frac{\omega_c}{2\pi l} \int_0^\infty dE \cos \frac{2\pi l E}{\omega_c} \frac{d}{dE}G''(\omega,E)\nonumber \\ &=& \int d\omega \frac{d n_f(\omega)}{d \omega} \frac{\omega_c}{2\pi l} \int_0^\infty dE \cos \frac{2\pi l E}{\omega_c} G''(\omega,E) \;.\nonumber \\
\end{eqnarray}
This expression can be treated by noting the derivative of the Fermi energy pins the $\omega$-integral to zero. This allows to treat the energy integration according to
\begin{eqnarray}
&&\int_0^\infty dE \cos \frac{2\pi l E}{\omega_c} G''(\omega,E)\nonumber \\ &=&\int_{-\mu}^\infty dE \cos \frac{2\pi l (E+\mu)}{\omega_c} G''(\omega,E-\mu) \nonumber \\ &\approx&\int_{-\infty}^\infty dE \cos \frac{2\pi l (E+\mu)}{\omega_c} G''(\omega,E-\mu) \nonumber \\&=&-\pi \left (\cos \frac{2 \pi l \mu}{\omega_c} \cos \frac{2 \pi l \omega}{\omega_c}- \sin \frac{2 \pi l \mu}{\omega_c} \sin \frac{2 \pi l \omega}{\omega_c}\right) e^{-\frac{l \pi}{\omega_c \tau}}\;.\nonumber \\
\end{eqnarray}
From there we can go on to solve the remaining integral. Since $\sin \frac{2 \pi l \omega}{\omega_c}$ is an odd function of $\omega$ and $\frac{d n_f(\omega)}{d\omega}$ even, the integral over $\sin \frac{2 \pi l \omega}{\omega_c}$ drops out leaving us with
\begin{eqnarray}
&& -\int d\omega \frac{d n_f(\omega)}{d \omega} \frac{\omega_c}{2 l}  \cos \frac{2 \pi l \mu}{\omega_c} \cos \frac{2 \pi l \omega}{\omega_c}e^{-\frac{\pi}{\omega_c \tau}}\nonumber \\&=& \frac{\omega_c}{2 l} \cos \frac{2 \pi l \mu}{\omega_c} e^{-\frac{\pi}{\omega_c \tau}} \frac{2\pi^2 l \frac{T}{\omega_c}}{\sinh \frac{2\pi^2 l T}{\omega_c}}\;.
\end{eqnarray}
We finally obtain (except for the diamagnetic contribution this can be compared to Ref.~\cite{Champel})
\begin{eqnarray}\label{eq:freeenergy}
\Omega&=&\Omega_0-\frac{\nu_0 \omega_c^2}{2\pi^2}\sum_{l=1}^\infty\frac{(-1)^l}{l^2} \left ( 1-\cos \frac{2\pi l \mu}{\omega_c}\frac{\lambda_l}{\sinh \lambda_l} e^{-\frac{l \pi }{\omega_c \tau}} \right) \nonumber \\ &=& \Omega_0+\frac{\nu_0 \omega_c^2}{24}+\frac{\nu_0 \omega_c^2}{2\pi^2}\sum_{l=1}^\infty\frac{(-1)^l}{l^2} \cos \frac{2\pi l \mu}{\omega_c}\frac{\lambda_l}{\sinh \lambda_l} e^{-\frac{l \pi }{\omega_c \tau}} \nonumber \\
\end{eqnarray}
where $\lambda_l=\frac{2\pi^2 l T}{\omega_c}$. 
In the limit $\omega_c \tau \ll 1$ we can approximate the grand potential as
\begin{eqnarray}
\Omega=\Omega_0+\frac{\nu_0 \omega_c^2}{24}-\frac{\nu_0 \omega_c^2}{2\pi^2}\cos \frac{2\pi \mu}{\omega_c}\frac{\lambda_1}{\sinh \lambda_1} e^{-\frac{ \pi }{\omega_c \tau}} \;.
\end{eqnarray}
One can easily obtain the $\frac{T}{\omega_c}\to 0 $ limit of this expression. This yields
\begin{eqnarray}
\Omega&=&\Omega_0+\frac{\nu_0 \omega_c^2}{24}-\frac{\nu_0 \omega_c^2}{2\pi^2}\cos \frac{2\pi \mu}{\omega_c}e^{-\frac{ \pi }{\omega_c \tau}} \nonumber \\ &=&\Omega_0+\frac{(eB)^2}{48 \pi m}-\frac{(eB)^2}{4\pi^3m}\cos \frac{2\pi \mu}{\omega_c}e^{-\frac{ \pi }{\omega_c \tau}} \;.
\end{eqnarray}
\section{The free energy of the oscillatory part of the photon system}\label{app:photon}

In the following we sketch the isolation of the temperature dependent part of the oscillatory part of the free energy. We start with an expression of the type
\begin{eqnarray}
f^A_{\rm{osc}}=\alpha \int_0^\Lambda q d q \int_{-\Lambda'}^{\Lambda'} d \omega n_b (\omega)\frac{\omega q^2}{A q^4 +\omega^2}
\end{eqnarray}
where $\Lambda\approx 2 k_f$ and $\Lambda'\approx E_f$. We first perform the integration with respect to $q$
\begin{eqnarray}
f^A_{\rm{osc}}=\frac{\alpha}{4A}\int_{-\Lambda'}^{\Lambda'} d \omega n_b (\omega) \omega \ln \left ( \frac{A \Lambda^4 +\omega^2}{\omega^2} \right)
\end{eqnarray}
and an integration by parts which leads to 
\begin{eqnarray}
f^A_{\rm{osc}} &=&   \frac{\alpha\left ( \Lambda' \right)^2}{8A} \ln \left ( 1+A\left ( \frac{\Lambda^2}{\Lambda'} \right)^2\right) \nonumber \\ &+&\frac{\alpha\Lambda^4}{8} \ln \left ( A \Lambda^4+\left( \Lambda' \right)^2 \right) \nonumber \\ &-&\frac{\alpha}{8A}\int_{-\Lambda'}^{\Lambda'} d \omega n_b' (\omega) \omega^2 \ln \left (1+\frac{A \Lambda^4}{\omega^2}\right) \nonumber \\ &-&\frac{\alpha \Lambda^4}{8}\int_{-\Lambda'}^{\Lambda'} d \omega n_b' (\omega)  \ln \left (\omega^2+A \Lambda^4\right) \;.
\end{eqnarray}
The first two terms can be discarded, since they have no temperature dependence. The last term will also not contribute a temperature dependent part to leading order in $\frac{T}{\Lambda'}$ and $\frac{T}{\Lambda^2}$. Proceeding with the remaining parts of the integral we obtain 
\begin{eqnarray}
f^A_{\rm{osc}} &\approx& -\frac{\alpha}{8 A} T^2 \ln \frac{\sqrt{A} \Lambda^2}{T} \int_{-\infty}^\infty d \omega \frac{\omega^2}{1-\cosh \omega} \nonumber \\ &+& \frac{\alpha}{8 A} T^2 \int_{-\infty}^\infty d \omega \frac{\omega^2 \ln |\omega|}{1-\cosh \omega} \nonumber \\ &=& \frac{\alpha}{6 A} \pi^2 T^2 \ln \frac{\sqrt{A} \Lambda^2}{T}-0.05880 \frac{\alpha}{A} T^2 \;.
\end{eqnarray}

\end{document}